\documentstyle[12pt,aps,fleqn]{revtex}
\input{psfig}

\newcommand{\beq}{\begin{eqnarray}}
\newcommand{\eeq}{\end{eqnarray}}
\begin{document}
\begin{center}
{\bf\LARGE Diffusion of a Deformable Body in a Random Flow}\\
\vspace{.7cm}
{\bf Gady Frenkel, Moshe Schwartz}\\
{\em Raymond and Beverly Sackler Faculty of Exact Sciences\\
School of Physics and Astronomy\\
Tel Aviv University, Ramat Aviv, 69978, Israel\\
}
\end{center}
\begin{abstract}
We consider a deformable body immersed in an incompressible liquid
that is randomly stirred. Sticking to physical situations in which the
body departs only slightly from its spherical shape, we calculate the
diffusion constant of the body. We give explicitly the dependence of
the diffusion constant on the velocity correlations in the liquid and
on the size of the body. We emphasize the particular case in which the
random velocity field follows from thermal agitation.\\

\end{abstract}
{\bf PACS:} 05.40.-a , 82.70.-y, 83.50.-v \\
{\bf Keywords:} drop diffusion, noise, deformable body, size effect,
deformation\\ 

%
%
%
%
%
%
%
Deformable bodies that are nearly spherical can be found in various
soft matter systems such as emulsions and complex fluids, where
deformable bodies are immersed in a host fluid. It is easy to
envisage a situations in which the host fluid is stirred randomly
\cite{libchaber,gang}. For example, mixing of bodies in
an emulsion by external mechanical vibrations (e.g., ultrasound).
Another example is thermal agitation of a complex
fluid. Hence, it
is interesting to investigate the dynamics of deformable bodies in
random flow. In this work we study the movement of a single deformable
body in a random velocity field, which is uncorrelated in time and
correlated in space in a general way. We focus on three aspects of
this subject: derivation of the explicit equation of motion for the
body's center, the Mean Squared Displacement (MSD) of the center and
the effect of the body's size on the diffusion constant.\\
Consider a single deformable body immersed in a host fluid.
The system is chosen to have the following characteristics:\\
\begin{enumerate}
\item The host fluid is incompressible,
$\vec{\nabla}\cdot\vec{v}=0$. Moreover, we assume that the Reynolds
number is small so that the Stokes approximation is applicable.
\item The body is characterized by an energy that depends on its
shape. The shape of minimum energy is a sphere. The
energy may be surface tension \cite{navot99}, Helfrich bending
energy \cite{helfrich76,lisy98}, etc.
The body's surface shape is described by the equation: 
$\psi(\vec{r})=0$ where $\psi(\vec{r})$ is a scalar three dimensional field.
\item A deformation of the body induces a force density on the host
fluid. As a result, the fluid's velocity is given by the sum of
$\vec{v}_{ext}$, which is caused by external sources, and
$\vec{v}_\psi$, which is induced by the body.
\item The external velocity, $\vec{v}_{ext}$, is random  
and is chosen to have zero average and known
correlations. It is convenient to define
the external velocity in terms of its spatial Fourier transform.
Since the fluid is incompressible, we will express the velocity in a
general form that has no longitudinal part:
\beq  \label{eq:vext}
v_{ext_i}(\vec{q}\ ) \equiv \sum_j \left( \delta_{ij}-\frac{q_i
    q_j}{q^2}\right) u_j(\vec{q}),
\eeq
where $\vec{u}$ is some general vector field.
Next, we define the correlations of the velocity by
the correlations 
of $\vec{u}$, 
\beq 
\left\langle u_l(\vec{q},t)\right\rangle = 0  \ \ and \nonumber
\eeq
\beq   \label{eq:fcorelation}
\left\langle u_l(\vec{q},t_1) u_m(\vec{p},t_2) \right\rangle =
\delta_{lm}\delta(\vec{q}+\vec{p})\phi(\xi q)\delta(t_2-t_1) ,
\eeq 
where $\delta_{lm}$ is the Kronecker delta, $\delta()$ is the Dirac delta function and $\xi$ is the velocity
correlation length.
\item The surface elements of the body are carried by the
host fluid \cite{schwartz90b}, i.e. each surface point moves according
to 
\beq   \label{eq:surfaceelement}
\dot{\vec{r}} = \vec{v}_{ext}(\vec{r}) + \vec{v}_\psi(\vec{r}).
\eeq
\item We assume that the external velocity is weak enough to cause
only minor shape fluctuations of the body.\\
\end{enumerate}
%
%
%
%
%

We will be interested in the following in the Mean Squared
Displacement (MSD) of the center. Since the body is deformable the
definition of its center is not unique. We will choose a specific
definition later. It turns out, however, that the value of the MSD at
long times does not depend on the specific choice, because for long
times the MSD (according to any definition) is much larger than the
size of the body.\\

Following the line of derivation of Schwartz and Edwards
\cite{schwartz90b,schwartz88}, equation (\ref{eq:surfaceelement}) may be
turned into a continuity equation for $\psi$,
\beq  \label{eq:continuity}
\frac{\partial \psi}{\partial t} + (\vec{v}_{ext} + \vec{v}_\psi)\cdot
\vec{\nabla}\psi = 0 .
\eeq
Consider a deformable body, carried by the host fluid in such a way
that at any instant it is nearly spherical. Its state can thus be
characterized by the position of its center, $\vec{r}_0(t)$, and a
deformation function $f(\Omega,t)$ that describes the shape by the
equation
\beq
\frac{\rho}{R} + f(\Omega,t) -1 = 0,
\eeq
where $\rho$ is the distance of the surface from the center in the direction of the
solid angle $\Omega$ and $R$ is the radius of the body when not
deformed. The deformation  function $f$ can be expanded in spherical
harmonics, $f(\Omega,t) = \sum_{l=0}^\infty \sum_{m=-l}^l
f_{l,m}(t)Y_{l,m}(\Omega)$. The center of the shape, $\vec{r}_0(t)$, is
defined as that point around which $f_{1m}(t)=0$ for $m=-1,0,1$. Parameterizing now
the gauge field $\psi$ as $\psi(\vec{r},t) = \frac{|\vec{r} - \vec{r}_0|}{R} 
+ f(\Omega,t) -1$, equation (\ref{eq:continuity}) leads to a linear
equation for each $f_{l,m}$,
\beq  \label{eq:flmdot}
\frac{\partial f_{lm}}{\partial t} + \lambda_l f_{lm}
+\frac{1}{R}[\hat{\rho}\cdot(\vec{v}_{ext} - \dot{\vec{r}}_0)]_{lm}=0  ,
\eeq
where $\hat{\rho}$ is a unit vector directed outwards from the center
in the direction of $\Omega$, and
\beq \label{eq:Qlm}
[\hat{\rho}\cdot(\vec{v}_{ext} - \dot{\vec{r}}_0)]_{lm} =
\int d\Omega \left\{
    \hat{\rho}\cdot\left[\vec{v}_{ext}(\vec{r}_0 + R(1-f)\hat{\rho}) -
      \dot{\vec{r}}_0\right]Y^*_{l,m}(\Omega)\right\}
\eeq
The eigenvalues $\lambda_l$'s characterize the decay rate of a slightly
deformed sphere into a sphere in the absence of the external
velocity. 
Note that the term $\lambda_l f_{lm}$  in Eq. (\ref{eq:flmdot})
results from the 
shape induced velocity, $\vec{v}_\psi$, in Eq. (\ref{eq:continuity})
\cite{schwartz88}.
Different physical systems are characterized by different
sets of $\lambda_l$'s. For example, Schwartz and Edwards
\cite{schwartz88} calculate $\lambda_l$ for a droplet with a constant surface
tension and equal 
viscosities inside and outside the droplet, Gang at al. \cite{gang} give
$\lambda_l$ for a droplet with a surface tension under
the assumption that the viscosity of its interior is much greater than
that of the surrounding fluid, Milner and Safran \cite{safran} do the
same for a surface 
controlled by bending energy and D\"orries and Foltin add to it
in-plane dissipation \cite{foltin}. In all of these cases, it is
obvious that the decay must depend only on $l$ because of the spherical
symmetry. Therefore, our treatment is general and applicable to a
large class of physical systems in which the decay rate
depends only on $l$.\\ 

Equation (\ref{eq:flmdot}) implies that in order that $f_{1,m}$
stays zero for all times we must have as an equation determining the
location of the center
\beq   \label{eq:motion1}
[\hat{\rho}\cdot(\vec{v}_{ext} - \dot{\vec{r}}_0)]_{1m} = 0 \ \ , \
m=-1,0,1 .
\eeq
%
%
%
%
%
%
%
%
%
For $l \neq 1$ it is clear that $\dot{\vec{r}}_0$ can be dropped from
the last term on the left hand side of
Eq. (\ref{eq:flmdot}). Therefore $f(\Omega,t)$ is linear in
$\vec{v}_{ext}$ (for long enough times the initial deformations have
already decayed). Consequently we can always drop, for small enough
$\vec{v}_{ext}$,  $f$ in the argument of
$\vec{v}_{ext}$ on the right hand side of Eq. (\ref{eq:Qlm}). The
result is decoupling of the deformation degrees of freedom from that
of the center of the sphere. The equation for the motion of the center
can thus be given, using linear combinations of $Y_{1,m}$, in vector form as
\beq
\int d\Omega \ \hat{\rho} \ 
\left(\hat{\rho}\cdot\dot{\vec{r}}_0 \right) =
\int d\Omega \ \hat{\rho} \ 
\left(\hat{\rho}\cdot\vec{v}_{ext}(\vec{r}_0 + R\hat{\rho})\right).
\eeq
Direct integration of the left hand side, Fourier decomposition of
$\vec{v}_{ext}(\vec{r}_0 + R\hat{\rho})$ on the right hand side and the
use of the partial waves decomposition \cite{bohm,landau}:
\begin{eqnarray}
e^{-i\vec{q}\cdot(R\hat{\rho})} = \sum_{l=0}^{\infty} \sum_{m=-l}^{l}
(-i)^l 4\pi j_l(qR)Y_{lm}^*(\Omega_q)Y_{lm}(\Omega) ,
\end{eqnarray}
where $\Omega_q$ is the solid angle in the $\vec{q}$ direction, yields
\begin{eqnarray} \label{eq:rDotAndA}
\dot{\vec{r}}_0 &=& 3 \int  d\vec{q} 
   e^{-i\vec{q}\cdot\vec{r}_0} \left( \frac{1}{3}j_0(qR) + 
   j_2(qR){\bf A}\right)\vec{v}_{ext}(\vec{q},t) ,
\end{eqnarray}
where $j_0$ and $j_2$ are spherical Bessel functions and the matrix ${\bf
  A}(\vec{q})$ is given by
\beq  \label{eq:matrixAij}
A_{ij}=-\frac{2}{3}\delta_{ij} +
\left(\delta_{ij} -\frac{q_i q_j}{q^2} \right).
\eeq
It may seem that ${\bf A}$ on the right hand side of
Eq. (\ref{eq:rDotAndA}) mixes directions. However, the bracketed
term in Eq. (\ref{eq:matrixAij}) is just a projection operator
on the transverse direction. The external velocity is incompressible
and hence already transverse. Consequently, this term acts as a unity
operator, $\delta_{ij}$ and  Eq. (\ref{eq:rDotAndA}) leads to 
\begin{eqnarray} \label{eq:velocity}
\dot{\vec{r}}_0 &=&  \int  d\vec{q} 
   e^{-i\vec{q}\cdot\vec{r}_0} \left( j_0(qR) + 
   j_2(qR)\right)\vec{v}_{ext}(\vec{q},t) .
\end{eqnarray}
Equation (\ref{eq:velocity}) is the explicit equation of motion for the
center of the body. In the limit $ R \rightarrow 0$ the 
approximation, $\dot{\vec{r}}_0 = \vec{v}_{ext}(\vec{r}_0,t)$ is obtained.
Note that this equation is general and describes the motion of the
center for any given (small enough) external velocity field.\\
%
%
%
%

%
Next, we calculate the MSD, $\langle \left( \Delta\vec{r}_0
  \right)^2 \rangle$, as a function of the elapsed time, $t$. The
starting point is the trivial relation 
\beq
\Delta \vec{r}_0(t) = \int_0^t \dot{\vec{r}}_0(t')dt'.
\eeq
Assuming the decomposition \cite{gady,brus88}
\beq
\left\langle v_{ext_i}(\vec{q}_1,t_1) v_{ext_j}(\vec{q}_2,t_2)
e^{-i\vec{q_1}\cdot\vec{r}_0(t_1)}e^{-i\vec{q_2}\cdot\vec{r}_0(t_2)}
\right.  \left.\right\rangle = \nonumber \\
\left\langle v_{ext_i}(\vec{q}_1,t_1) \right.\left.v_{ext_j}(\vec{q}_2,t_2) \right\rangle \left\langle
e^{-i\vec{q_1}\cdot\vec{r}_0(t_1)}e^{-i\vec{q_2}\cdot\vec{r}_0(t_2)}
\right\rangle
\eeq
and using
equations (\ref{eq:vext}), (\ref{eq:fcorelation}) and (\ref{eq:velocity})
the MSD is
\beq \label{eq:MSDintegral}
\left\langle \left( \Delta\vec{r}_0(t) \right)^2 \right\rangle =
 8\pi \int_0^\infty q^2dq
 \phi (q)\left(j_0(qR)+j_2(qR)\right)^2 t \equiv 3 D t.
\eeq
%
%
%
%
%
%
%
We expect the diffusion coefficient to depend on the ratio $\gamma =
\frac{R}{\xi}$ in the following way. As $\gamma$ increases the diffusion coefficient 
decreases. This is due to the fact that as $\gamma$ 
increases, different regions of the surface become less correlated and
move in different directions. In the limit  $\gamma \rightarrow
\infty$, the movement of the center ceases and the MSD
is always zero. In the limit  $\gamma \rightarrow 0$, the bracketed
Bessel term in Eq. (\ref{eq:MSDintegral}) can be replaced by unity. A close inspection of the
derivation reveals that this limit produces the same MSD equation as
the approximated equation $\dot{\vec{r}}_0
=\vec{v}_{ext}(\vec{r}_0)$ that is accurate
in the limit of infinite correlation length, or point particles.\\
%
%
We turn to evaluate the dependence of the diffusion coefficient on $R$
and $\xi$ in the above limiting cases.
Consider the correlation function $\phi (q) = C\ (q\xi)^\alpha
g(q\xi)$, where $g$ is a cutoff function and $g(0)=C_0 > 0$. 
The diffusion constant is given by
\beq
D = \frac{8\pi C \xi^\alpha}{3 R^{3+\alpha}}\int_0^\infty du \ u^{2+\alpha}
 g(\frac{\xi}{R}u) [j_0(u)+j_2(u)]^2 .
\eeq
In the limit $R/\xi \rightarrow\infty$ we distinguish between two
cases: $\alpha<1$ and $\alpha>1$. Since the large $u$ dependence of
$j_0(u) + j_2(u)$ is proportional to $cos(u)/u^2$ we find that
\beq  \label{eq:Dinfinity}
D \propto \left\{ \begin{array}{ll}
\frac{C \xi^\alpha}{R^{3+\alpha}} & $for $ \alpha<1\\
\frac{C\xi}{R^4} & $for $\alpha>1.\\
\end{array} \right.
\eeq
In the opposite limit $R/\xi \rightarrow 0$ we find that regardless of
$\alpha$
\beq \label{eq:Dzero}
D \propto \frac{C}{\xi^3}.
\eeq
(Note here that we have written the power law dependence of $\phi(q)$
as $q^\alpha \xi^\alpha$ but having other dimensional constants in the
model may make $C$ depend on $\xi$, so that eqs. (\ref{eq:Dinfinity}) and
(\ref{eq:Dzero}) may be considered only as equations that yield the
dependence of $D$ on the radius $R$).\\
It is interesting to consider the case where the fluctuations in the
velocity field are due to thermal agitation. 
%
General considerations show that the correlations of the external
velocity must be, in this case, of the form
\beq \label{eq:22}
\left\langle v_{ext_i}(\vec{q},t_1) v_{ext_j}(\vec{p},t_2) \right\rangle =
\left(\delta_{ij} -\frac{q_iq_j}{q^2}\right)\delta(\vec{q}+\vec{p})\delta(t_2-t_1)\frac{c}{q^2} ,
\eeq
where $c$ is a dimensional constant (Strictly speaking the above is
valid for $qa>1$, where $a$ is the inter-particle distance). 
We use these correlations with Eq. (\ref{eq:MSDintegral}) and obtain
\beq
D=\frac{8 \pi^2 c}{5R}.
\eeq
Dimensional analysis reveals that $c$ must be proportional to  
$K_B T / \eta$ (with a dimensionless proportionality
constant). A detailed calculation yields a proportionality constant
equal to $(2\pi)^{-3}$. We do not represent here the detailed
calculation, because it yields a result identical to that obtained in
the past \cite{schwartz91,hadamard} using two totally different
approaches (that we
wanted to check by a more direct calculation. Actually, Hadamard calculated the mobility that can be
related to the diffusion constant by the Einstein relation). 
Therefore, for $R$ larger than the inter-particle distance in the liquid,
\beq
D = \frac{K_B T}{5\pi \eta R}.
\eeq

\begin{figure}
\centerline{\psfig{figure=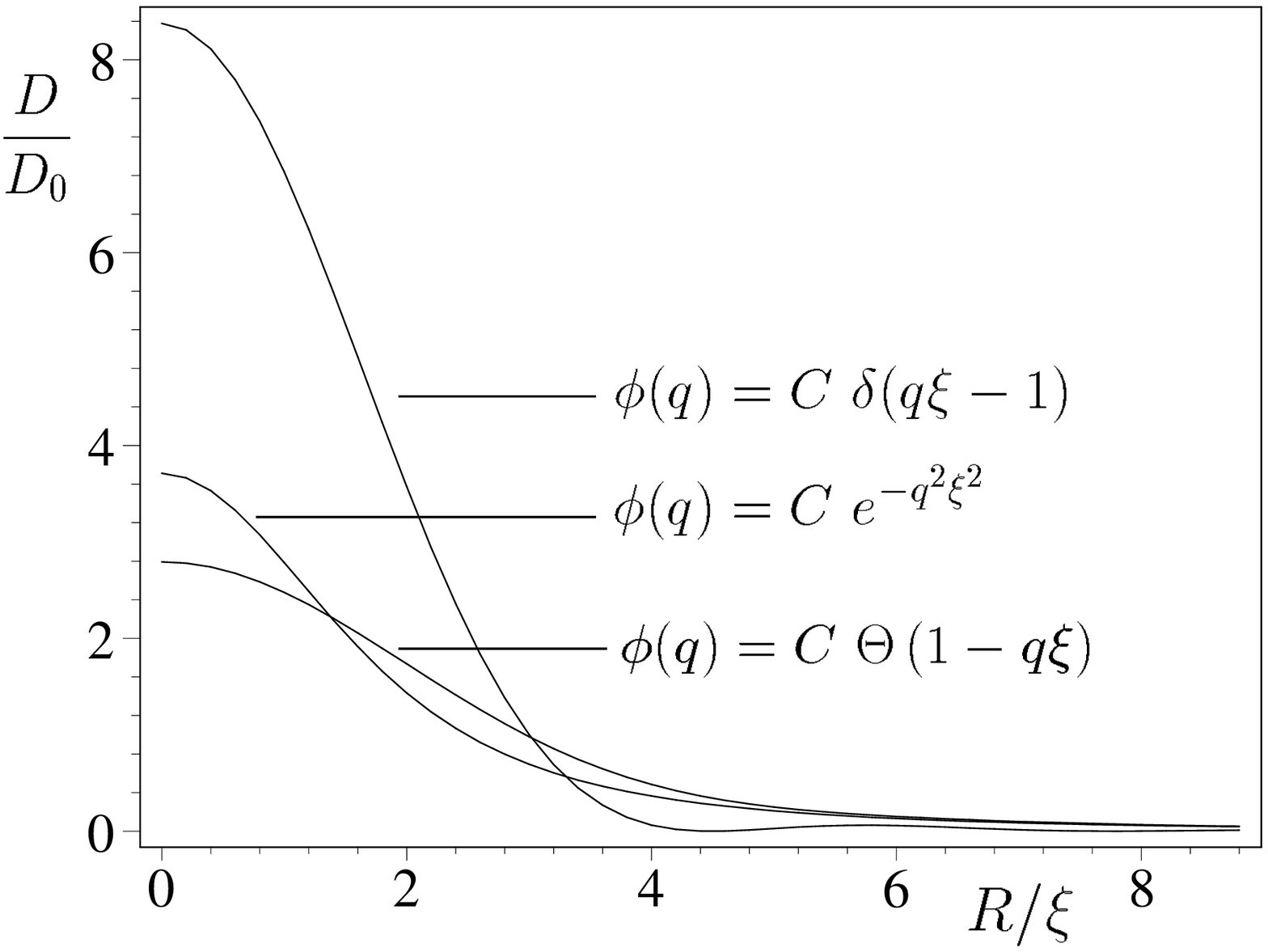,width=18cm ,height=14cm,clip=}}
\caption{}

\label{fig:wnoise1}
\end{figure}

\begin{figure} 
\centerline{\psfig{figure=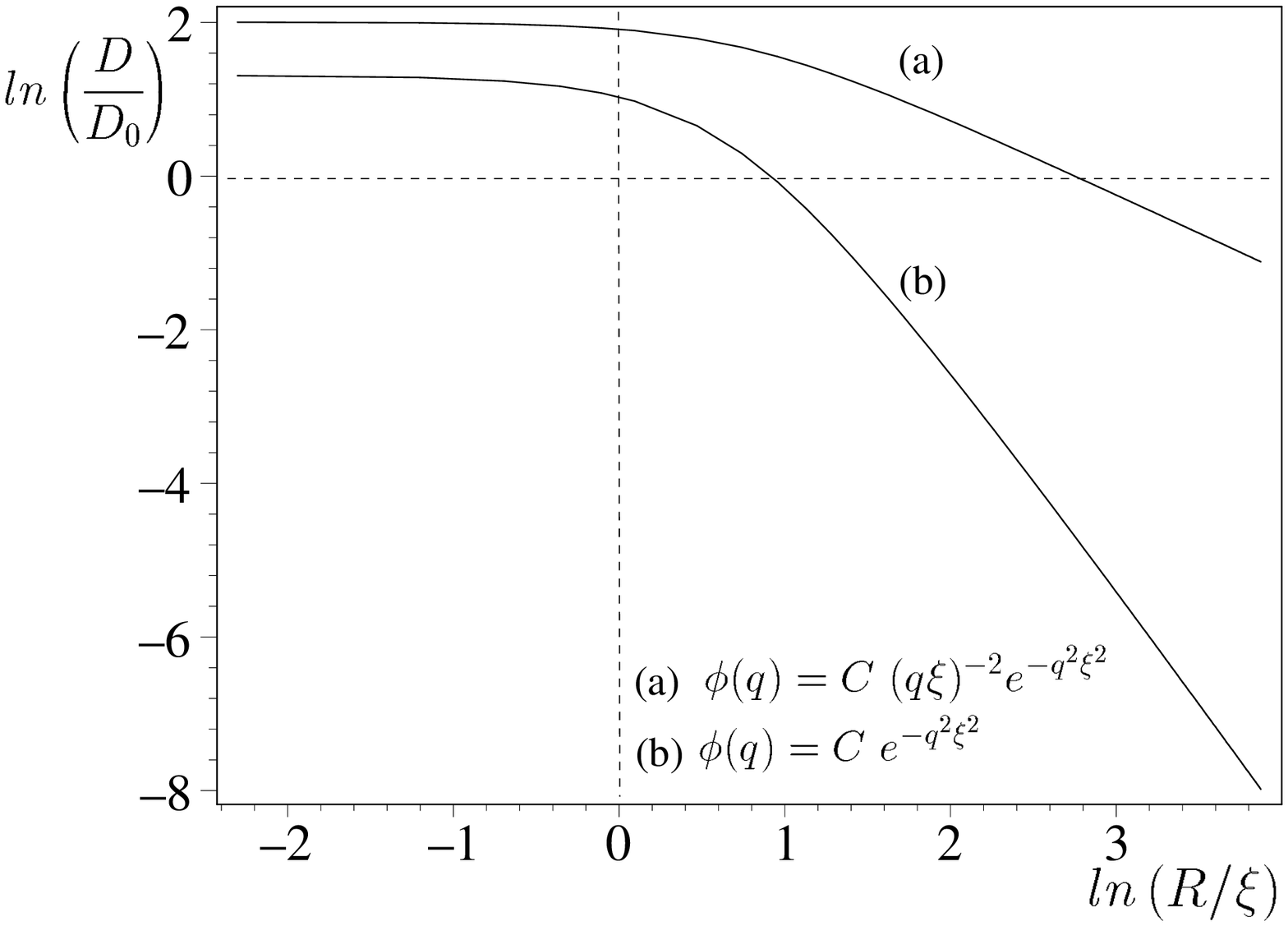,width=18cm ,height=14cm,clip=}}
\caption{}

\label{Rdependence}
\end{figure}

Note that this result, for a liquid membrane that has liquid inside
as well as outside, is different from the well known Stokes result for a hard
sphere and from the result
for a polymer subjected to thermal fluctuations \cite{doi} (The
difference is in the prefactors).\\
Fig. \ref{fig:wnoise1} depicts $D$ for typical correlation functions while Fig. \ref{Rdependence} depicts the dependence of the diffusion coefficient on
$R,\xi$, for two noise realizations. The first is just a cutoff
function, $\phi (q) = C\ exp(-q^2\xi^2)$. The second corresponds to
the temperature driven randomness, $\phi (q) = C\
(q\xi)^{-2} exp(-q^2\xi^2)$. We keep $\xi$ 
constant and vary $R$. There are two distinct regimes: For $ R/\xi <1
$ the diffusion coefficient is not sensitive to $R$, while for
$R/\xi>1$ the slope of the graph turns towards -3 for the cutoff function and
towards -1 for the temperature driven randomness.\\
\newpage
{\bf Figure captions}\\

{\bf Fig. 1}\\
The dependence of the diffusion coefficient $D$  on the ratio of
the radius $R$ of the sphere to the correlation length $\xi$, for typical
correlation functions. The axis are presented in non-dimensional
units, where $D_0=C/\xi^3$.\\

{\bf Fig. 2}\\
The dependence of the diffusion coefficient on $R$. We keep
$\xi$ constant and vary $R$. For negative $ln(R/\xi)$ there is no R
dependence, while for positive values $D$ is proportional to
$R^\mu$ where $\mu$ tends to (a) -1 for thermal agitation and (b) -3 for a simple cutoff function.
\newpage
\newpage

\end{document}